\pdfoutput=1
\documentclass[prb,twocolumn,showpacs,amsmath,amssymb,floatfix,superscriptaddress]{revtex4-1}

\usepackage{graphicx}
\usepackage{dcolumn}
\usepackage{bm}
\usepackage{hyperref}

\begin{document}

\title{Nonequilibrium spin texture within a thin layer below the surface of current-carrying topological insulator Bi$_2$Se$_3$: A first-principles quantum transport study}

\author{Po-Hao Chang}
\affiliation{Department of Physics and Astronomy, University of Delaware, Newark, DE 19716-2570, USA}
\author{Troels Markussen}
\affiliation{QuantumWise A/S, Fruebjergvej 3, Box 4, DK-2100 Copenhagen, Denmark}
\author{S{\o}ren Smidstrup}
\affiliation{QuantumWise A/S, Fruebjergvej 3, Box 4, DK-2100 Copenhagen, Denmark}
\author{Kurt Stokbro}
\affiliation{QuantumWise A/S, Fruebjergvej 3, Box 4, DK-2100 Copenhagen, Denmark}
\author{Branislav K. Nikoli\'{c}}
\email{bnikolic@udel.edu}
\affiliation{Department of Physics and Astronomy, University of Delaware, Newark, DE 19716-2570, USA}

\begin{abstract}
We predict that unpolarized charge current injected into a ballistic thin film of prototypical topological insulator (TI) Bi$_2$Se$_3$  
will generate a {\it noncollinear spin texture} $\mathbf{S}(\mathbf{r})$ on its surface. Furthermore, the nonequilibrium spin texture will extend 
into $\simeq 2$ nm thick layer below the TI surfaces due to penetration of evanescent wavefunctions from the metallic surfaces into the bulk of TI. Averaging $\mathbf{S}(\mathbf{r})$ over few \AA{} along the longitudinal direction defined by the current flow reveals large component pointing in the transverse direction. In addition, we find an order of magnitude smaller out-of-plane component when the direction of injected current with respect to Bi and Se atoms probes the largest hexagonal warping of the Dirac-cone dispersion on TI surface. Our analysis is based on an extension of the nonequilibrium Green functions combined with density functional theory (NEGF+DFT) to situations involving noncollinear spins and spin-orbit coupling. We also demonstrate how DFT calculations with properly optimized local orbital basis set can precisely match putatively more accurate calculations with plane-wave basis set for the supercell of Bi$_2$Se$_3$.
\end{abstract}

\maketitle

The newly discovered three-dimensional topological insulator (3D TIs) materials possess a usual band gap in the bulk while also hosting metallic surfaces. The low-energy quasiparticles on these surfaces behave as massless Dirac fermions whose spins are locked to their momenta due to strong spin-orbit coupling (SOC).~\cite{Hasan2010} Such spin-momentum locking is viewed as a resource for spintronic applications.~\cite{Pesin2012} For example, very recent experiments~\cite{Mellnik2014} have demonstrated magnetization dynamics of a single ferromagnetic metallic (FM) overlayer deposited on the surface of 3D TIs due to current-induced SO torques. Another recent experiment~\cite{Shiomi2014} has detected spin-to-charge conversion~\cite{Mahfouzi2014a,Shen2014} when precessing magnetization of the FM overlayer pumps pure spin current into the metallic surface of 3D TIs.

The microscopic mechanism behind these phenomena can be traced to the so-called Edelstein effect (EE), originally predicted~\cite{Edelstein1990} for a {\em diffusive} two-dimensional  electron gas (2DEG) with the Rashba SOC~\cite{Winkler2003} and observed much later experimentally.~\cite{Kato2004} In the EE in 2DEG,  longitudinal unpolarized charge current flowing along the $x$-axis drives a homogeneous nonequilibrium spin density $\mathbf{S}=(0,S_y,0)$ pointing in the transverse direction. The diffusive metallic surface of TIs also exhibits EE where a current-driven spin density $\mathbf{S}$ is substantially enhanced~\cite{Pesin2012} (by a factor $\hbar v_F/\alpha_R \gg 1$, with $v_F$ being the Fermi velocity in TI and $\alpha_R$ is the strength~\cite{Winkler2003} of the Rashba SOC in 2DEG). This enhancement can be explained by the spin-momentum locking along the single Fermi circle,~\cite{Hasan2010} formed in $k$-space at the intersection of the Dirac cone energy-momentum dispersion and the Fermi energy plane, in contrast to spin-momentum locking along the two circles~\cite{Winkler2003} in the case of Rashba 2DEG which counter the effect of each other. This has motivated recent experiments~\cite{Li2014} probing $\mathbf{S}$ directly in three-terminal geometry where nonmagnetic electrodes inject unpolarized charge current into a TI, while a third FM contact deposited in the middle of the top surface of the TI film detects a voltage signal when a non-zero $\mathbf{S}$ is induced. These setups quantify the projection of $\mathbf{S}$ onto the magnetization of the third FM contact.

\begin{figure}
\includegraphics[scale=0.1,angle=0]{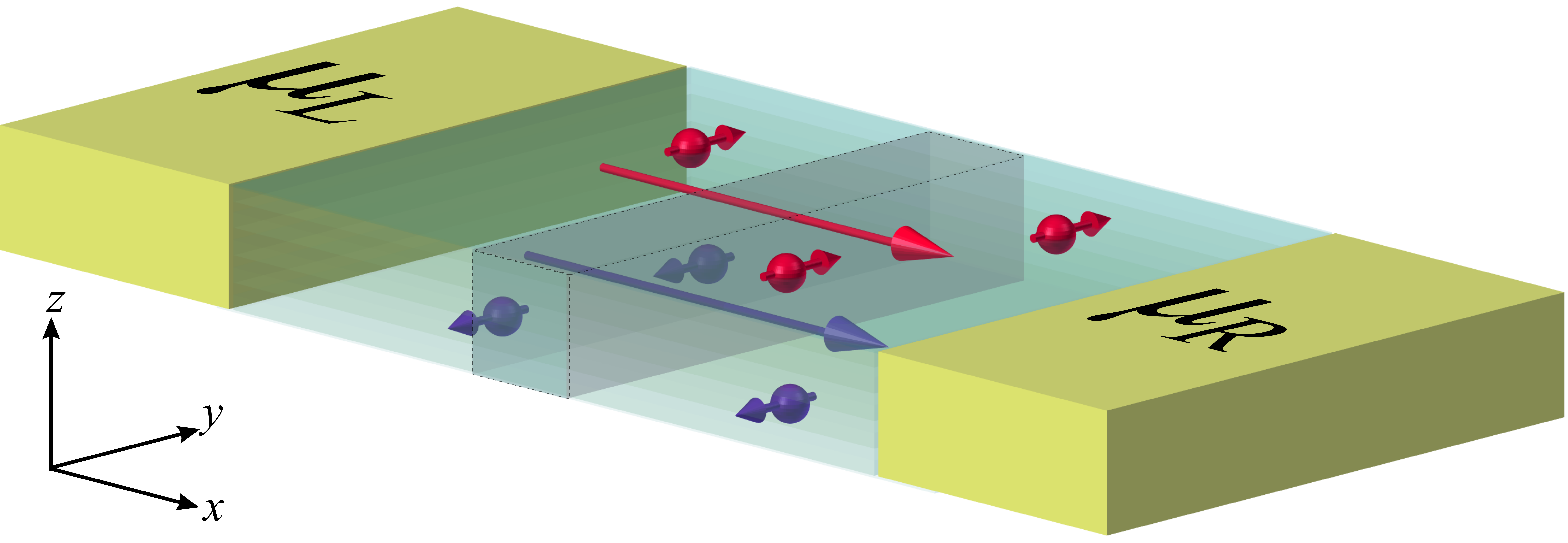}
\caption{Schematic view of a two-terminal setup where a thin film of Bi$_2$Se$_3$ is attached to two macroscopic reservoirs
biased by the electrochemical potential difference $eV_b = \mu_L - \mu_R$. The {\em clean} Bi$_2$Se$_3$ film is infinite along the $x$-axis
(i.e., the direction of transport) and the $y$-axis, while its thickness along the $z$-axis is chosen as 5 QLs. The shaded cell of length \mbox{$d_x \simeq 5$ \AA{}} defines volume for averaging $\mathbf{S}(\mathbf{r})$ from Fig.~\ref{fig:fig2}, which is then plotted in Fig.~\ref{fig:fig3} over the corresponding cross section of thin film within the $yz$-plane.}
\label{fig:fig1}
\end{figure}
\begin{figure*}
\includegraphics[scale=0.45,angle=0]{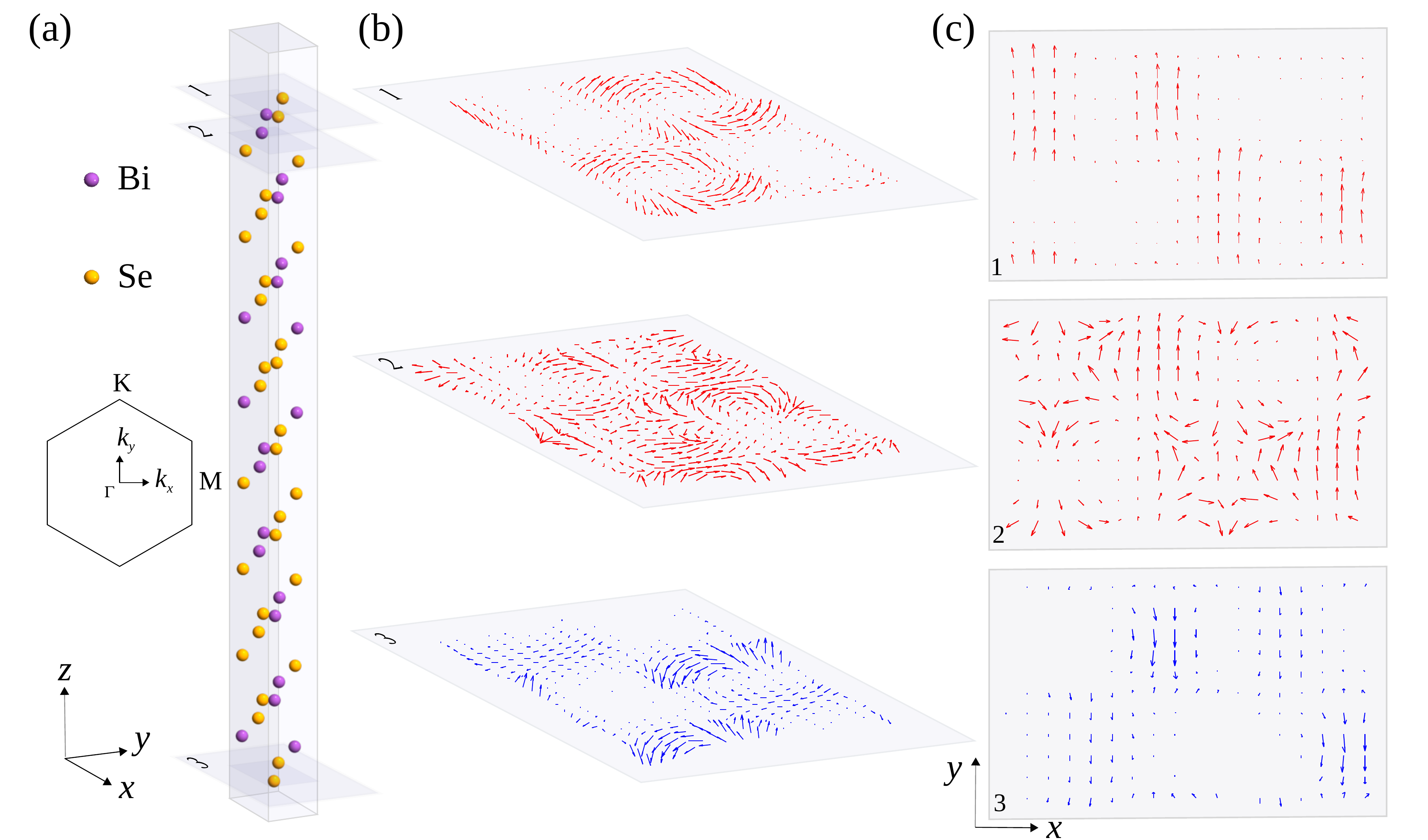}
\caption{(a) The arrangement of Bi and Se atoms within a supercell of Bi$_2$Se$_3$ thin film  of thickness 5 QLs. The inset in panel (a) shows BZ in the $k_x$-$k_y$ plane at $k_z$ = 0 with special $k$-points $\Gamma$, $M$, and $K$ marked. (b) The vector field of nonequilibrium $\mathbf{S}(\mathbf{r})$ within selected planes shown in (a), generated by injection of unpolarized charge current along the $x$-axis (see also Fig.~\ref{fig:fig1}). The planes 1 and 3 correspond to the top and bottom metallic surfaces of Bi$_2$Se$_3$ thin film, while plane 2 resides in the bulk at a distance \mbox{$d \approx 0.164$ nm} away from plane 1. (c) The vector fields in (b) projected onto each of the selected planes in (a). The real space grid of $\mathbf{r}$-points in panels (b) and (c) has spacing $\simeq 0.4$ \AA{}.}
\label{fig:fig2}
\end{figure*}

However, this picture of EE on the surface of TI is based on simplistic model Hamiltonians.~\cite{Mellnik2014,Pesin2012} Here we employ first-principles quantum transport approach to analyze microscopic  details, over \mbox{$\lesssim 1$ \AA{}} length scale, of current-driven $\mathbf{S}(\mathbf{r})$ in the two-terminal {\em ballistic} thin film geometry hosting realistic TI material, as illustrated in Fig.~\ref{fig:fig1}. We choose Bi$_2$Se$_3$ as the prototypical TI material---with its single Dirac cone in the surface band structure, relatively large bulk band gap,
and Dirac point (DP) inside the gap [see Fig.~\ref{fig:fig4}(a)]---on
which many recent experiments probing EE directly~\cite{Li2014} or
indirectly~\cite{Mellnik2014,Shiomi2014} have been performed. The
central region of the device in Fig.~\ref{fig:fig1}, which has
length \mbox{$L_x = 21.5$ \AA{}} along the $x$-axis and infinite width
along the $y$-axis, is attached to two semi-infinite electrodes made
of the same material. The electrodes are assumed to terminate at
infinity into macroscopic Fermi liquid reservoirs where electrons are
thermalized to acquire electrochemical potential $\mu_L$ in the left
reservoir and $\mu_R$ in the right one. The Hamiltonian of the central
region and the electrodes is obtained from the noncollinear density functional
theory (ncDFT), implemented by us in ATK package~\cite{atk}, using optimized pseudo-atomic localized basis functions~\cite{Ozaki2003} and SOC introduced via the total-angular-momentum-dependent pseudopotentials.~\cite{Theurich2001}  The transport properties of the system in Fig.~\ref{fig:fig1} 
are computed using the nonequilibrium Green function (NEGF) formalism,~\cite{Stefanucci2013} so that our approach 
represents an extension of the widely used NEGF+DFT framework~\cite{Taylor2001} to transport problems involving 
noncollinear spins and SOC.

In the simplest picture---based on  effective
Hamiltonian  $\hat{H}_\mathrm{TI} = v_F
(\hat{\bm \sigma} \times \hat{\mathbf{p}}) \cdot \mathbf{e}_z$
($\hat{\bm \sigma}$ is the vector of the Pauli matrices;
$\hat{\mathbf{p}}$ is the momentum operator; and $\mathbf{e}_z$ is the
unit vector along the $z$-axis in Fig.~\ref{fig:fig1}) describing
massless Dirac electrons on the metallic surfaces of TIs---the spin and momentum of
electronic eigenstates  are orthogonal to each other along the single
Fermi circle. This generates net homogeneous $\mathbf{S} = (0, S_y, 0)$ after an applied electric field $E_x$ shifts the Fermi
circle~\cite{Pesin2012,Mellnik2014,Misawa2011} along the momentum
parallel to $E_x$. Such manifestation of EE persists in ballistic
samples as well~\cite{Modak2012,Chang2014a} where there is no electric
field within the TI but instead one applies bias voltage $eV_b = \mu_L - \mu_R$  
to inject a current into the TI, as illustrated in Fig~\ref{fig:fig1}. The relations $S_y \propto E_x$ or $S_y \propto V_b$ describing EE in the diffusive or ballistic transport regimes, respectively, are allowed only in nonequilibrium since in equilibrium $\mathbf{S}$ changes sign under time reversal, and, therefore, has to vanish (assuming absence of magnetic field).

\begin{figure}
\includegraphics[scale=0.55,angle=0]{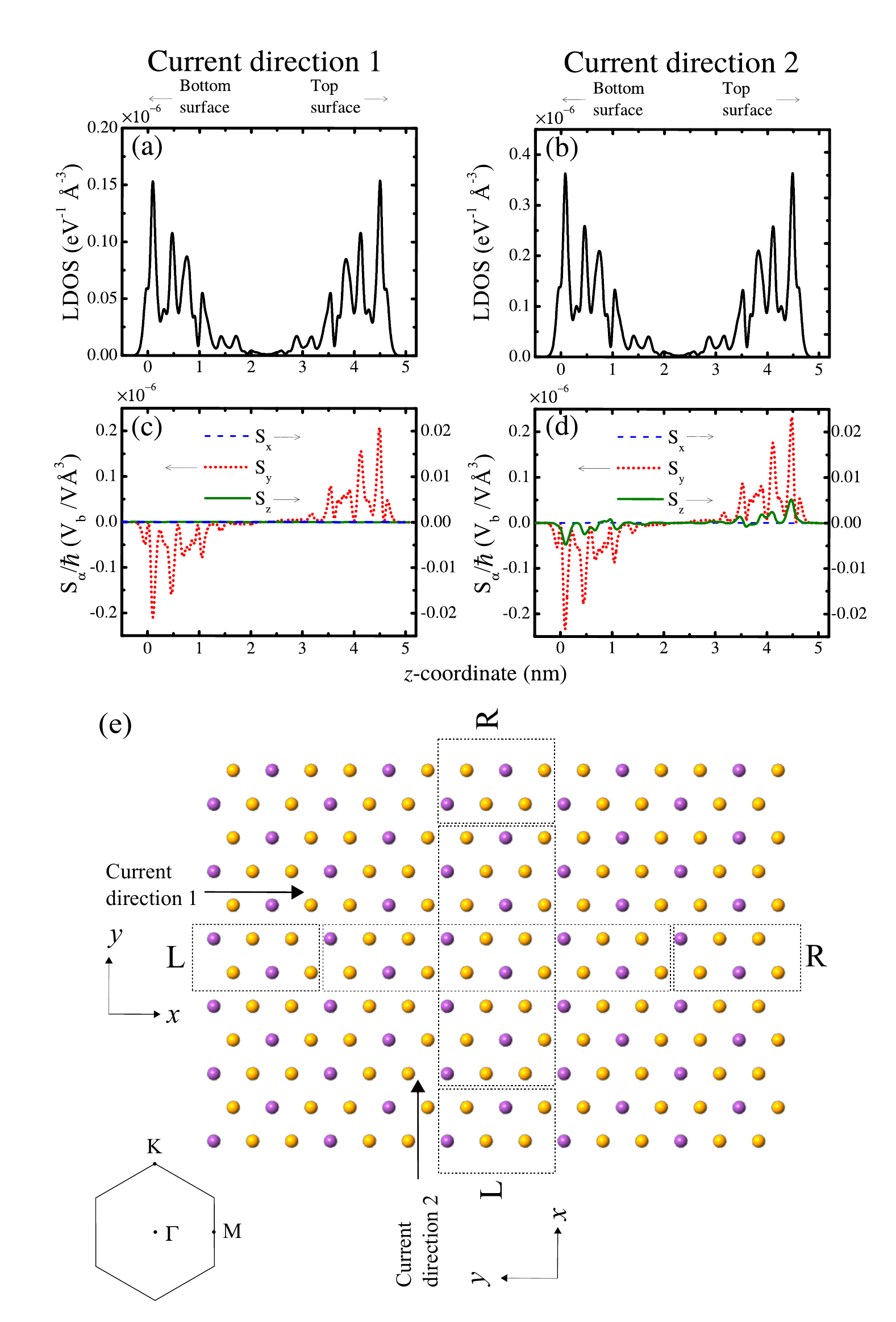}
\caption{(a),(b) The spatial profile of LDOS at $E_F$
over the cross section of the cell denoted in
Fig.~\ref{fig:fig1}. (c),(d) The spatial profile of the components of
$\mathbf{S}(\mathbf{r})$, obtained by averaging its texture plotted in
Fig.~\ref{fig:fig2}, over the cell of Bi$_2$Se$_3$ thin film marked in 
Fig.~\ref{fig:fig1}. The direction of injected charge current for the results in panels (a) and (c), or the 
results in panels (b) and (d), is denoted in panel (e) relative to the orientation of the lattice 
of Bi and Se atoms. The bottom surface of Bi$_2$Se$_3$ is located at \mbox{$z=0$ nm}, and the top TI surface is located at \mbox{$z \approx 4.56$ nm}.}
\label{fig:fig3}
\end{figure}

\begin{figure}
\includegraphics[scale=0.35,angle=0]{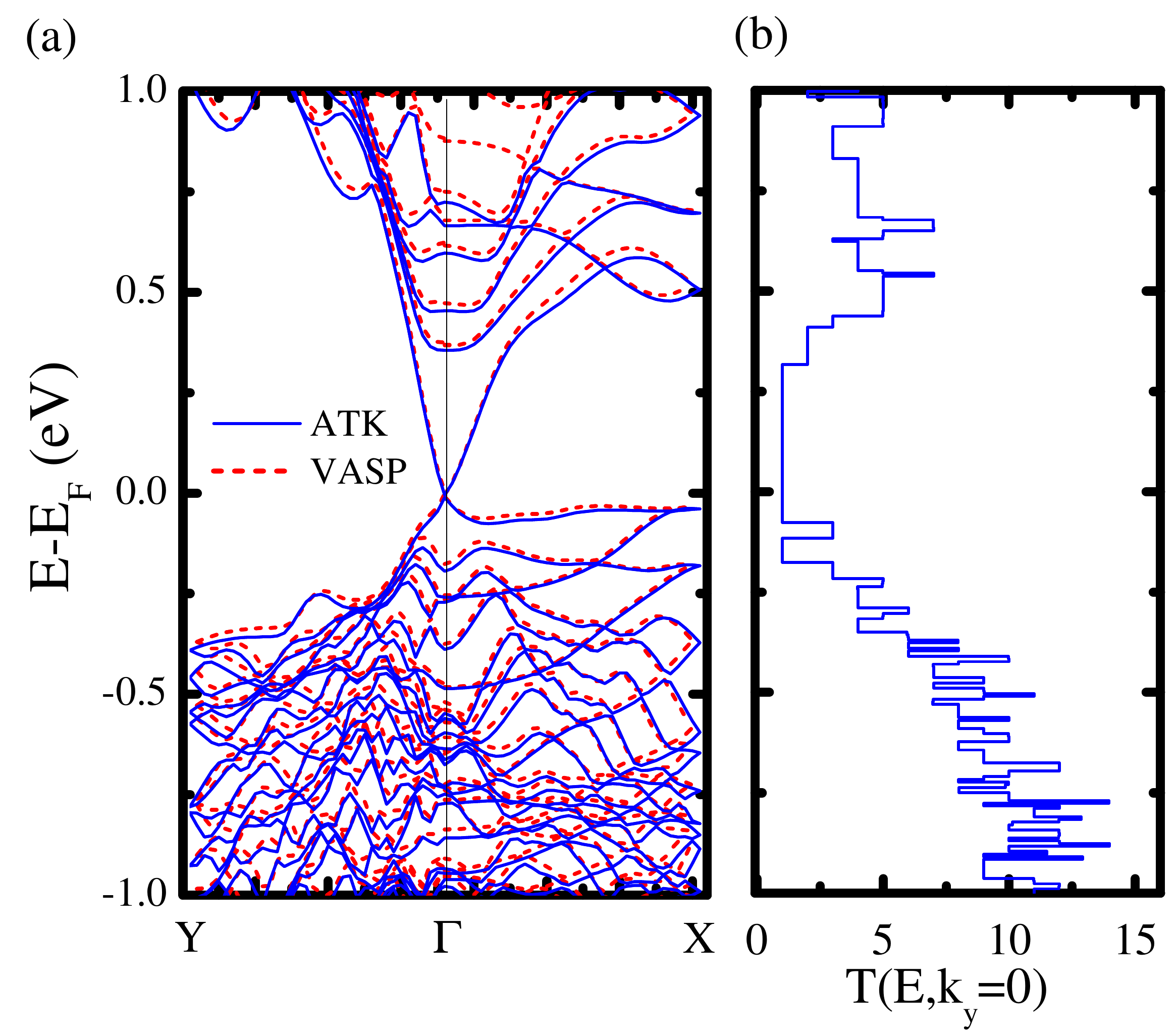}
\caption{(a) The electronic band structure computed for a supercell of Bi$_2$Se$_3$ film shown in Fig.~\ref{fig:fig2}(a) using LCAO~\cite{Ozaki2003} ncDFT  implemented in ATK~\cite{atk} package. This is compared with the electronic band structure obtained using PW ncDFT implemented in VASP package.~\cite{Kresse1993} (b) Zero-bias transmission function of Bi$_2$Se$_3$ thin film in the two-terminal geometry of Fig.~\ref{fig:fig1}, for electrons injected along the \mbox{$\Gamma$--$M$} direction $(k_y=0)$ in the inset of Fig.~\ref{fig:fig2}(a), computed using NEGF+ncDFT formalism implemented in ATK package.}
\label{fig:fig4}
\end{figure}

This simplistic picture can be contrasted with our principal results 
in Figs.~\ref{fig:fig2} and ~\ref{fig:fig3}. When a small (ensuring linear-response transport
regime) $V_b$ is applied between the reservoirs in
Fig.~\ref{fig:fig1}, the unpolarized charge current injected into
Bi$_2$Se$_3$ thin film generates a nonequilibrium  
$\mathbf{S}(\mathbf{r})$ whose complex {\it noncollinear} texture within three planes selected in
Fig.~\ref{fig:fig2}(a) is plotted in Figs.~\ref{fig:fig2}(b) and
~\ref{fig:fig2}(c). For the visualization we use real-space grid for $\mathbf{r}$ whose
spacing is $\simeq 0.4$ \AA{}. Furthermore, Figs.~\ref{fig:fig2} and ~\ref{fig:fig3} demonstrate that nonequilibrium 
spin texture $\mathbf{S}(\mathbf{r})$ will appear not only on the TI surface, but
also within \mbox{$\simeq 2$ nm} thick layer of its bulk just below  
the top and bottom surfaces. This feature is explained in
Figs.~\ref{fig:fig3}(a) and ~\ref{fig:fig3}(b) showing spatial profile
of the local density of states (LDOS) at the Fermi energy $E_F$ over
the cell depicted in Fig.~\ref{fig:fig1}. The non-zero LDOS and the
corresponding $\mathbf{S}(\mathbf{r})$ in the bulk of the TI thin film
stem from evanescent wavefunctions which originate from the top and
bottom metallic surfaces and penetrate into the energy gap of the
insulating bulk. The Bi$_2$Se$_3$ is a strongly anisotropic material
composed of quintuple layers (QLs) of Bi and Se atoms, illustrated in
Fig.~\ref{fig:fig2}(a), where one QL consists of three Se layers
strongly bonded to two Bi layers in between. For Bi$_2$Se$_3$ film
thinner than 5 QLs, the evanescent wavefunctions from the top and
bottom metallic surface can overlap to create a
minigap~\cite{Yazyev2010,Park2010} at the DP. We select the thickness
of Bi$_2$Se$_3$ to be 5 QLs along the $z$-axis in Fig.~\ref{fig:fig1},
which ensures that the LDOS in Figs.~\ref{fig:fig3}(a) and ~\ref{fig:fig3}(b) goes to zero on the plane half way between the top and bottom surfaces of the TI thin film.

Upon averaging nonequilibrium $\mathbf{S}(\mathbf{r})$ over a \mbox{$d_x \simeq 5$ \AA{}} long cell depicted in Fig.~\ref{fig:fig1}, we obtain spatial profiles in Figs.~\ref{fig:fig3}(c) and ~\ref{fig:fig3}(d) which show that $S_y$ is the largest component independently of the direction of incoming electrons. An order of magnitude smaller $S_z$ component shown in Fig.~\ref{fig:fig3}(d) appears for electrons incoming along current direction 2 marked in panel (e). This is in accord with experiments in equilibrium where spin- and angle-resolved photoemission spectroscopy~\cite{Pan2011} finds largest out-of-plane spin component along the corresponding direction in the 2D Brillouin zone (BZ). This is due to hexagonal warping of the Dirac cone surface band structure, as confirmed by DFT calculations~\cite{Yazyev2010,Zhao2011} finding that equilibrium expectation value of spin in the eigenstates of Bi$_2$Se$_3$ surfaces tilts out of the 2D BZ. Thus, Fig.~\ref{fig:fig3} offers a novel prescription for probing hexagonal warping even close to DP via transport measurements where charge current is injected in different directions relative to the orientation of the lattice of Bi and Se atoms.


We now explain the technical details of our calculations. The extension of DFT to the case of spin-polarized systems is formally derived 
in terms of total electron density $n(\mathbf{r})$ and vector  magnetization density $\mathbf{m}(\mathbf{r})$. In the collinear DFT,
$\mathbf{m}(\mathbf{r})$ points in the same direction at all points in space, which is insufficient to study magnetic systems where the
direction of the local magnetization is not constrained to a particular axis or systems  governed by SOC. In 
ncDFT,~\cite{Capelle2001} the exchange-correlation (XC) functional $E_\mathrm{XC}[n(\mathbf{r}),\mathbf{m}(\mathbf{r})]$ depends on $\mathbf{m}(\mathbf{r})$ pointing in arbitrary directions. The local density approximation(LDA) and most often
employed version of generalized gradient approximation (GGA), implemented also by us in ATK,~\cite{atk} make
additional approximations~\cite{Capelle2001} that lead to the XC magnetic field \mbox{$\mathbf{B}_\mathrm{XC}(\mathbf{r}) = \delta E_\mathrm{XC}[n(\mathbf{r}),\mathbf{m}(\mathbf{r})]/\delta \mathbf{m}(\mathbf{r})$} being parallel everywhere to $\mathbf{m}(\mathbf{r})$.

The single-particle spin-dependent Kohn-Sham (KS) Hamiltonian in  ncDFT takes the form
\begin{equation}\label{eq:ks}
   \hat{H}_\mathrm{KS} = -\frac{\hbar^2\nabla^2}{2m}  + V_\mathrm{H}({\bf r}) + V_{\rm XC}({\bf r}) + V_{\rm ext}({\bf r}) - {\bm \sigma} \cdot \mathbf{B}_\mathrm{XC}(\mathbf{r}),
\end{equation}
where $V_H({\bf r})$, $V_{\rm ext}({\bf r})$ and $V_\mathrm{XC}({\bf r})=E_\mathrm{XC}[n(\mathbf{r}),\mathbf{m}(\mathbf{r})]/\delta n(\mathbf{r})$ are the Hartree, external and XC potential, respectively. Diagonalization of   $\hat{H}_\mathrm{KS}$ proceeds by
approximating the Hilbert space of all single-electron eigenfunctions
with a finite set of basis functions. A popular basis set is
plane-waves (PWs), where varying only one parameter (the energy
cutoff) allows one to improve the basis systematically. Linear
combination of atomic orbitals (LCAO) basis sets require more tuning,
however, they simplify the  NEGF calculations~\cite{Taylor2001} where one has to spatially separate system into the central region and semi-infinite electrodes, as illustrated in Fig.~\ref{fig:fig1}.

Since the pioneering screening~\cite{Zhang2009} of candidate TI materials via ncDFT calculations, their electronic band structure 
has most often been calculated~\cite{Yazyev2010,Park2010} using PW ncDFT 
with electron-core interactions described via projector augmented wave (PAW) method.~\cite{Blochl1994} 
In Fig.~\ref{fig:fig4}(a) we demonstrate that such calculations, performed by VASP 
package,~\cite{Kresse1993} can be accurately reproduced by pseudopotential-based LCAO ncDFT implemented 
in ATK.~\cite{atk} The supercell considered in both calculations is shown in Fig.~\ref{fig:fig2}(a), which includes 5 QLs terminated by Se atomic layer on both the top and bottom surface, as well as \mbox{$7.5$ \AA{}} thick vacuum layer above and below these Se atomic layers. 

We note that previous attempts~\cite{Zhao2011} to apply pseudopotential-based LCAO ncDFT to Bi$_2$Se$_3$ have yielded either poor accuracy of its electronic band  structure (e.g., compare our Fig.~\ref{fig:fig4}(a) with Fig. 1 in Ref.~\onlinecite{Zhao2011}) or have required intricate fine tuning.~\cite{Rivera2014} Therefore, we provide here a complete recipe for the proper usage of LCAO ncDFT to reproduce Fig.~\ref{fig:fig4}(a). In ATK calculations in Fig.~\ref{fig:fig4}(a), the electron-core interactions are described by  norm-conserving pseudopotentials. The pseudopotentials
are obtained by mapping the solution of the Dirac equation, which
naturally includes SOC,~\cite{Theurich2001} to non-relativistic
pseudopotential, $V_\mathrm{PS} = V_\mathrm{L} + V_\mathrm{NL}^{1/2} +
V_\mathrm{NL}^{-1/2}$, with local contribution $V_\mathrm{L}$ and
non-local contributions $V_\mathrm{NL}$ from the total angular
momentum $j=l+1/2$ and $j=l-1/2$. The non-local terms are expanded in
terms of SO projector functions, $V_\mathrm{NL}^{\pm 1/2}
= \sum_{l,\xi,\alpha,\beta} \nu_{l\pm 1/2,\xi} P_{\alpha\beta}^{l \pm
1/2, \xi}$, where  $\nu_{l \pm 1/2,\xi}$ are normalization constants
and the indices $\alpha,\beta$ denote the possible spin orientations
$(\uparrow,\downarrow)$. We use Perdew-Burke-Ernzerhof (PBE) parametrization of GGA for the XC functional  
and a LCAO basis set $\{ \phi_i \}$ generated by the  OpenMX
package,~\cite{openmx,Ozaki2003} which consists of {\tt s2p2d1} orbitals on
Se atoms and {\tt s2p2d2} on Bi atoms. These  pseudoatomic orbitals were
generated by a confinement scheme~\cite{Ozaki2003} with the cutoff
radius 7.0 a.u. and 8.0 a.u. for Se and Bi atoms,  respectively. The
energy mesh cutoff for the real-space grid is chosen as 75.0 Hartree. In VASP calculations~\cite{Kresse1993} in Fig.~\ref{fig:fig4}(a), the electron-core interactions  are described by PAW method,~\cite{Blochl1994} and we employ PBE GGA for the XC functional. The cutoff energy for the PW basis set is \mbox{350 eV}. In both ATK and VASP calculations we employ \mbox{$11 \times 11 \times 1$} $k$-point mesh within Monkhorst-Pack scheme for the BZ integration.

The eigenstates $|\Psi_n \rangle$ of the KS Hamiltonian in Eq.~\eqref{eq:ks} make it possible to construct the equilibrium density matrix ${\bm \rho}_\mathrm{eq} = \sum_n |\Psi_n \rangle \langle \Psi_n| f(E)$ for electrons at $\mu_L = \mu_R$ and temperature $T$
described by the Fermi distribution function $f(E)$. The local
electron and magnetization density, as the central variables of ncDFT,
are obtained from \mbox{$n(\mathbf{r})
= \langle \mathbf{r}|\mathrm{Tr}_\mathrm{spin}[{\bm \rho}_\mathrm{eq}]|\mathbf{r} \rangle$}
and \mbox{$\mathbf{m}(\mathbf{r})
= \langle \mathbf{r}| \mathrm{Tr}_\mathrm{spin}
[{\bm \rho}_\mathrm{eq} {\bm \sigma}]| \mathbf{r} \rangle$}, where the
trace is taken over the spin Hilbert space. 

In steady-state nonequilibrium due to dc current flowing between the left and right reservoirs in Fig.~\ref{fig:fig1}, we construct the nonequilibrium density matrix~\cite{Mahfouzi2013} ${\bm \rho}_\mathrm{neq}$ using NEGFs:
\begin{equation}\label{eq:noneqrho}
{\bm \rho}_\mathrm{neq} = \frac{1}{2\pi i} \int\limits_{-\infty}^{+\infty} dE\, \mathbf{G}^<(E) - {\bm \rho}_\mathrm{eq}.
\end{equation}
This yields $\mathbf{S}(\mathbf{r})= \frac{\hbar}{2} \langle \mathbf{r}| \mathrm{Tr}_\mathrm{spin} [{\bm \rho}_\mathrm{neq} {\bm \sigma}] |\mathbf{r}\rangle$ plotted in Figs.~\ref{fig:fig2} and ~\ref{fig:fig3}. The NEGF formalism~\cite{Stefanucci2013} for steady-state transport operates with two central quantities---the retarded GF, $\mathbf{G}(E)$, and the lesser GF, $\mathbf{G}^<(E)$---which describe the density of available quantum states and how electrons occupy those states, respectively. In the absence of inelastic processes, these are given by \mbox{$\mathbf{G} =[E \mathbf{O} - \mathbf{H}_\mathrm{KS} - {\bm \Sigma}_L - {\bm \Sigma}_R]$} and \mbox{$\mathbf{G}^< = i\mathbf{G}[f_L \mathbf{\Gamma}_L + f_R \mathbf{\Gamma}_R] \mathbf{G}^\dagger$}. Here the self-energies ${\bm \Sigma}_{L,R}$ are due to semi-infinite electrodes, $f_{L,R}=f(E-\mu_{L,R})$ and \mbox{${\bm \Gamma}_{L,R}=i ({\bm \Sigma}_{L,R} - {\bm \Sigma}_{L,R}^\dagger)$} is the level broadening matrix. For the chosen LCAO basis set, the Hamiltonian matrix $\mathbf{H}_\mathrm{KS}$ is composed of elements  $\langle \phi_i|\hat{H}_\mathrm{KS}|\phi_j\rangle$ and the overlap matrix $\mathbf{O}$ is composed of elements $\langle \phi_i|\phi_j \rangle$. In the linear-response transport regime considered here, Eq.~\eqref{eq:noneqrho} can be expanded~\cite{Mahfouzi2013} to first order in bias voltage $V_b$. Since $\mathbf{S}(\mathbf{r})$ is zero in equilibrium (because of assumed absence of external magnetic field), the linear-response density matrix can be  simplified  to,~\cite{Mahfouzi2013} ${\bm \rho}_\mathrm{neq} = \frac{eV_b}{2\pi} \int\limits_{-\infty}^{+\infty} dE\, \mathbf{G} {\bm \Gamma}_L  \mathbf{G}^\dagger \left( - \frac{\partial f}{\partial E} \right)$. Otherwise, the gauge-invariant form of ${\bm \rho}_\mathrm{neq}$ requires additional terms~\cite{Mahfouzi2013} to properly remove the equilibrium expectation value of a considered physical quantity.

The retarded GF also allows us to obtain the transmission function of
the device in
Fig.~\ref{fig:fig1}, \mbox{$T(E,k_y)=\mathrm{Tr}[{\bm \Gamma}_R \mathbf{G}
{\bm \Gamma}_L \mathbf{G}^\dagger]$}, which depends on energy and
transverse momentum $k_y$ due to assumed periodicity in the
$y$-direction. The total transmission function $T(E)$ is obtained by integrating over $k_y$, which determines the linear-response conductance via the Landauer formula, $G = \frac{e^2}{h} \int dE\, T(E) \left( - \frac{\partial f}{\partial E} \right)$. We confirm in Fig.~\ref{fig:fig4}(b) that  $T(E,k_y=0)=2$ for $E$ within the bulk gap shown in Fig.~\ref{fig:fig4}(a) because only one doubly degenerate helical conducting channel is open for transport in that energy range~\cite{Wang2012c} for injected electrons with momentum along the \mbox{$\Gamma$--M} direction ($k_y = 0$).

In conclusion, using NEGF+ncDFT framework  implemented  by
us in ATK package,~\cite{atk} we computed a nonequilibrium spin
texture  $\mathbf{S}(\mathbf{r})$ within a thin film of
current-carrying Bi$_2$Se$_3$ TI material. The non-zero texture
appears on the TI metallic top and bottom surfaces, as well as within bulk
layers of thickness \mbox{$\simeq 2.0$ nm} below the surfaces
that effectively dope the bulk by evanescent wavefunctions. The spin
texture is  noncollinear and complex on length scales \mbox{$\lesssim
1$ \AA{}}. Upon averaging it over a few \AA{} we find a simpler pattern---with 
either $\mathbf{S} = (0,S_y,0)$, or $\mathbf{S} = (0,S'_y,S'_z)$ where $S'_y/S'_z \gg 1$---depending on the direction of injected current with respect to orientation of the lattice of Bi and Se atoms. Such dependency offers a novel probe, via  electronic transport measurements,~\cite{Li2014} of the hexagonal warping of the Dirac cone surface band structure. For the envisaged spintronic applications of TIs, it is essential to understand how $\mathbf{S}(\mathbf{r})$ changes due to finite bias voltage or self-consistent coupling~\cite{Semenov2014} to magnetization of a ferromagnetic (metal or insulator) overlayer, which we relegate to future studies.

\begin{acknowledgments}
P.-H. C. and B. K. N. were supported by NSF Grant No. ECCS 1509094. The supercomputing time was provided by XSEDE, which is
supported by NSF Grant No. ACI-1053575. QuantumWise acknowledges support from the Danish Innovation Fund Grant No. 79-2013-1: 
``Nano-scale design tools for the semiconductor industry.''
\end{acknowledgments}




\end{document}